\documentclass[twocolumn,aps,prb,superscriptaddress,showpacs,amsmath,amssymb]{revtex4}

\usepackage{graphicx}
\usepackage{dcolumn}
\usepackage{bm}

\begin{document}


\title{Incommensurate interactions and non-conventional
spin-Peierls transition in TiOBr}

\author{Sander van Smaalen}
\email{smash@uni-bayreuth.de}
\homepage{http://www.uni-bayreuth.de/departments/crystal/}
\affiliation{Laboratory of Crystallography, University of Bayreuth,
95440 Bayreuth, Germany}
\author{Lukas Palatinus}
\affiliation{Laboratory of Crystallography, University of Bayreuth,
95440 Bayreuth, Germany}
\affiliation{Institute of Physics, Czech Academy of Sciences,
162 53 Prague, Czechia}%
\author{Andreas Sch\"onleber}
\affiliation{Laboratory of Crystallography, University of Bayreuth,
95440 Bayreuth, Germany}

\date{\today}

\begin{abstract}

Temperature-dependent x-ray diffraction of the low-dimensional
spin 1/2 quantum magnet TiOBr shows that the phase
transition at $T_{c2}=47.1 (4)$ K corresponds to the development
of an incommensurate superstructure.
Below $T_{c1}=26.8 \pm 0.3$ K the incommensurate modulation locks
in into a two-fold superstructure similar to the low-temperature
spin-Peierls state of TiOCl.
Frustration between intra- and interchain interations within
the spin-Peierls scenario,
and competition between two-dimensional magnetic order and
one-dimensional spin-Peierls order
are discussed as possible sources of the incommensurability.
\end{abstract}

\maketitle

\section{INTRODUCTION}

Recently, TiOCl was proposed to be a quasi-one-dimensional (1D)
$S = 1/2$ quantum spin system, that develops a spin-Peierls
state at low temperatures.\cite{seidela2003}
The spin-Peierls state is now well established by the temperature
dependence of the magnetic susceptibility ($\chi_m$), that
is zero below the phase transition at $T_{c1}=67$ K,
the observation by NMR of two independent Ti
atoms below $T_{c1}$, the two-fold crystallographic
superstructure below $T_{c1}$ and electronic band-structure
calculations.\cite{seidela2003,imait2003,shazm2005,sahadasgupta2004}
The atomic displacements in the superstructure
as well as the calculated band structure, with the single valence
electron of Ti$^{3+}$ occupying the $d_{xy}$ orbital,
indicate that the spin-Peierls state is formed on the chains of
Ti atoms parallel to $\mathbf{b}$ via direct exchange interactions.
\cite{seidela2003,shazm2005,sahadasgupta2004}

Although the properties of the low-temperature phase of TiOCl
are those of a true spin-Peierls system, TiOCl is not a
conventional spin-Peierls compound, because the phase
transition at $T_{c1}$ is first-order.
The temperature dependencies of
$\chi_m$, electron spin resonance (ESR), nuclear magnetic
resonance (NMR), specific heat ($C_p$) and x-ray diffraction
have shown that a second-order phase transition occurs
at $T_{c2}=91$ K.
\cite{seidela2003,kataevv2003,imait2003,hembergerj2005,shazm2005}
The 1D character of the magnetic interactions was also supported
by the temperature dependencies of optical reflectivity and
angle-resolved photoelectron spectroscopy (ARPES),
\cite{caimig2004a,hoinkism2004,ruckampr2005a}
although it was suggested that on cooling from room temperature,
a crossover from two-dimensional (2D) towards 1D interactions
occurs.\cite{caimig2004a,lemmensp2004a}
The nature of the state above $T_{c1}$ is not understood yet.
Orbital, spin and structural fluctuations have been proposed
to be responsible for the properties of TiOCl.
\cite{imait2003,gracol2004,pisanil2005,hembergerj2005}
However, R\"uckamp \textit{et al.}\cite{ruckampr2005a} suggested
that orbital fluctuations can be ruled out.

TiOBr and TiOCl crystallize
in the FeOCl structure type.\cite{schaferh1958,vonschnering1972}
Physical properties of both compounds are similar,
with the two transition temperatures scaled down towards
$T_{c1} = 27$ K and $T_{c2} = 47$ K in TiOBr.
\cite{caimig2004b,katoc2005,lemmensp2005,sasakit2005,ruckampr2005a}
The two-fold superstructures below $T_{c1}$ are similar
in TiOCl and TiOBr, suggesting a spin-Peierls state for
TiOBr too.\cite{palatinusl2005a}
In the present contribution we report the discovery of
incommensurate satellite reflections in x-ray diffraction
of TiOBr at temperatures $T$ with $T_{c1} < T < T_{c2}$.
Complete crystal structures are presented, but the data do
not allow to distinguish between a one-dimensional and a
two-dimensional incommensurate modulation wave.
These two models provide two possible interpretations for
the understanding of the interatomic interactions in TiOBr
and TiOCl.

\section{EXPERIMENTAL}

Single crystals of TiOBr were prepared by gas transport
reaction.\cite{schaferh1958,palatinusl2005a}
A single crystal of dimensions
$0.27 \times 0.13 \times 0.002$ mm$^{3}$
was glued on a carbon fiber that was attached to a closed-cycle
helium cryostat mounted on a four-circle Huber diffractometer.
Single-crystal x-ray diffraction with synchrotron radiation
was measured at beam-line D3 of Hasylab (DESY, Hamburg), employing
monochromatized radiation of wavelength $0.5000 (1)$ {\AA}
and a point detector.

\begin{figure}
{\includegraphics[width=3.9cm]{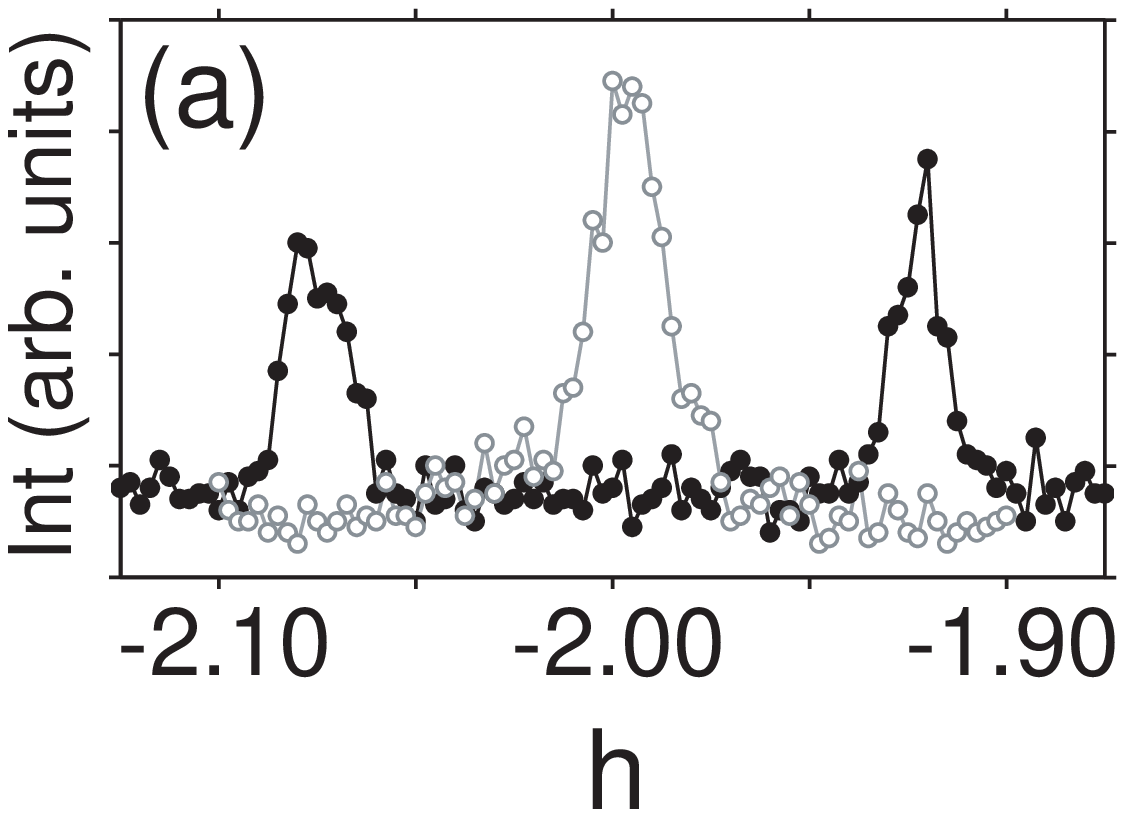}}
{\includegraphics[width=3.9cm]{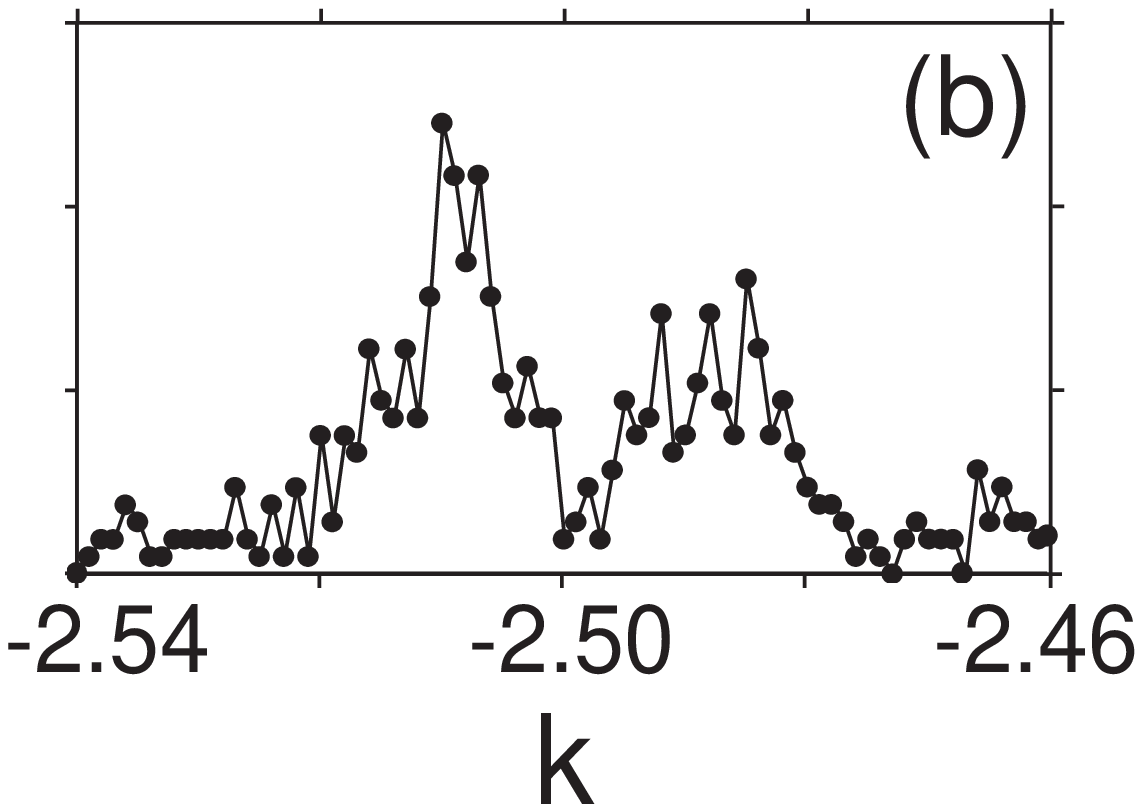}}
\caption{\label{f-qscans}
(a) Q-scans along $\mathbf{a}^*$ centered on $({-2},  {-3.5},  {-1})$
at $T=17.5$ K (open circles) and $T = 37$ K (full circles).
(b) Q-scans along $\mathbf{b}^*$ centered on $({0.075},  {-2.5},  {-1})$
at $T = 35$ K.
Lines are a guide for the eye.
Step-widths of the scans were $0.0025$.
Intensities in (a) are higher than those in (b).
}
\end{figure}

\begin{figure}
{\includegraphics[width=3.9cm]{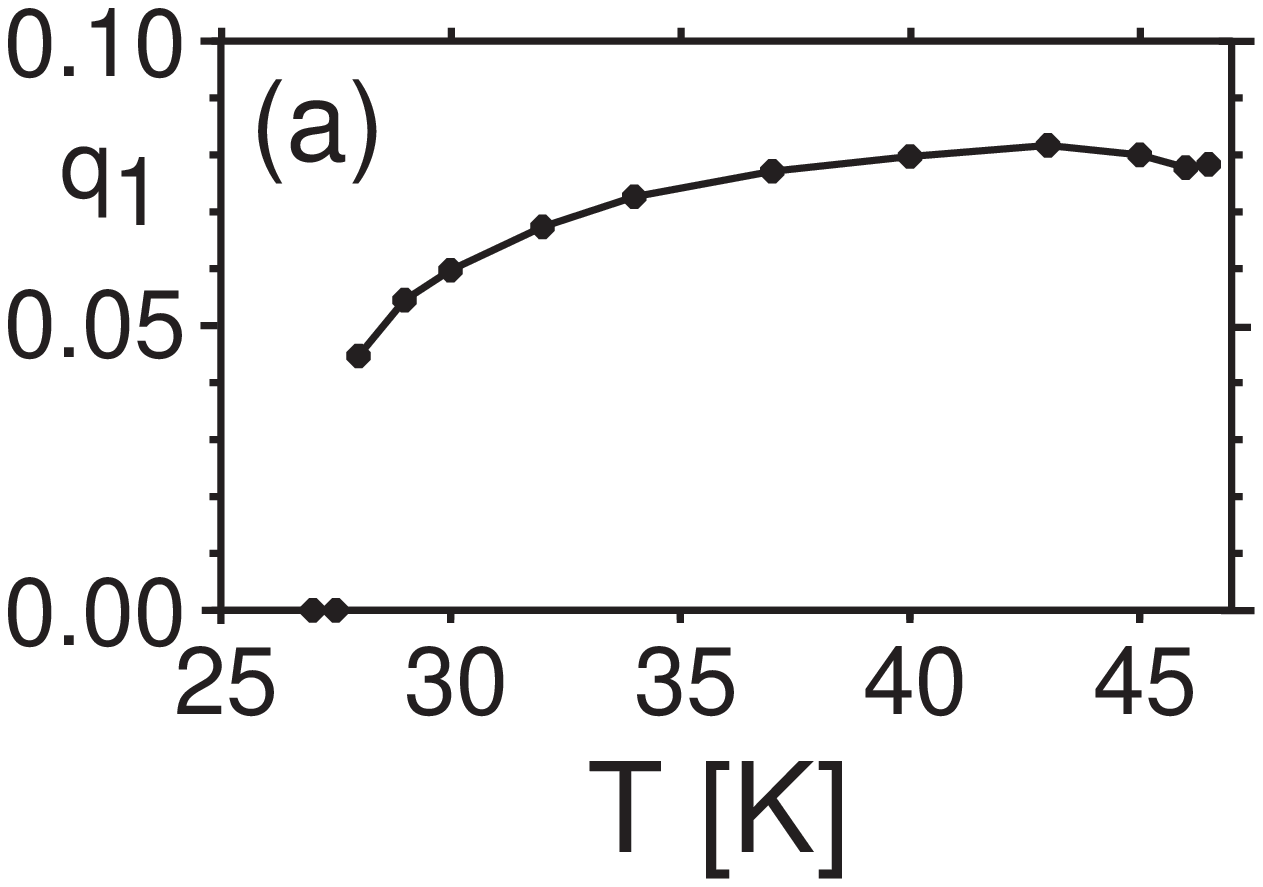}}
{\includegraphics[width=3.9cm]{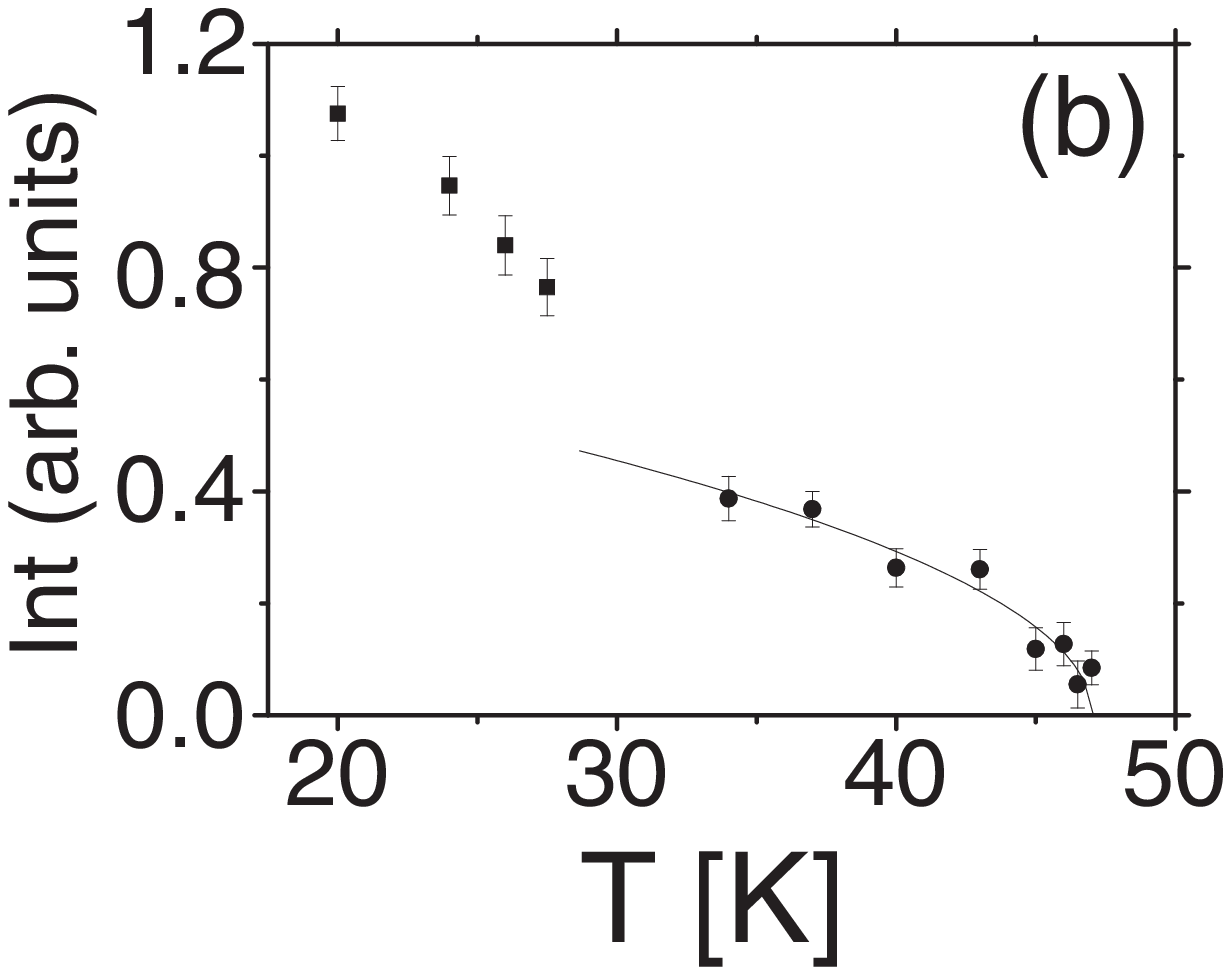}}
\caption{\label{f-sats_int+pos}
(a) Temperature dependence of $q_1$ in $\mathbf{q} = (q_1, 0.5, 0)$.
$q_1$ was obtained as half the distance between the
two maxima in $q$-scans, like those given in Fig. \ref{f-qscans}a.
(b) Temperature dependencies of the intensities of the satellite
reflections $({-2},  {-3.5},  {-1})$ (squares)
and $({2+q_1},  {-3.5},  {-1})$ (circles).
The line represents a fit with the function
$I(T) = I_0\: \sqrt{1-T/T_{c2}}$ with $T_{c2}=47.0(6)$ K.
Similar plots lead to
$T_{c2}=47.1(6)$ K for $({-2-q_1}, {-3.5}, {-1})$,
$T_{c2}=46.8(8)$ K for $({1+q_1},  {-3.5},  {-2})$ and
$T_{c2}=48(2)$ K for $({1-q_1},  {-3.5},  {-2})$.
}
\end{figure}

Diffraction at room-temperature confirmed the FeOCl structure
type.\cite{palatinusl2005a}
The temperature dependence of the component $q_1$ of the
modulation wavevector $\mathbf{q} = (q_1, 0.5, 0)$
was determined from $q$-scans along $\mathbf{a}^*$ centered on
the positions $({-2},  {-3.5},  {-1})$ and $(1,  {-3.5},  {-2})$.
Up to $T_{c1} = 27$ K a single peak was found at $q_1 = 0$,
while for $T_{c1} < T < T_{c2} = 47$ K two peaks appeared in
each scan, at positions $\pm q_1$ (Fig. \ref{f-qscans}a).
Above $T_{c2}$ any diffraction at these positions had disappeared.
These results show that below $T_{c1}$ TiOBr has a two-fold
superstructure, while in the intermediate phase TiOBr is
incommensurately modulated.
The component $q_1$ of the modulation wavevector
was found to continuously decrease on decreasing temperature,
and it jumps to zero at $T_{c1}$ (Fig. \ref{f-sats_int+pos}a).
This result corroborates the incommensurate character
of the modulation in the intermediate phase, and it shows the
first-order character of the transition at $T_{c1}$.
The transition temperature was determined from
Fig. \ref{f-sats_int+pos}a as $T_{c1} = 26.8\pm0.3$ K.

The integrated intensities of reflections were measured by
$\omega$-scans centered at the expected reflection positions.
In this way, the temperature dependencies were measured of the
intensities of two
commensurate satellites (below 27 K) and of four incommensurate
satellites (Fig. \ref{f-sats_int+pos}b).
Due to an experimental error, reliable intensities were
not obtained for temperatures $27.5$ K $\le T \le 32$ K.
The intensities $I(T)$ above $T_{c1}$ could be fitted
by a function $I(T) = I_0\: \sqrt{1-T/T_{c2}}$,
that provided consistent estimates for the transition
temperature from all four reflections,
with an average value of $T_{c2} = 47.1 (4)$ K.
The gradual loss of intensity of the incommensurate satellites
on approaching $T_{c2}$ is an indication for the second-order
character of the phase transition at this temperature.

Incited by the observation, that some reflections were broader
than others in the $\omega$-scans of the data collection (see below),
the possibility of a second incommensurate component of the
modulation was investigated,
then resulting in a modulation
wavevector $\mathbf{q} = (q_1, {0.5+\delta}, 0)$.
Evidence for a splitting along $\mathbf{b}^*$ was obtained
from $q$-scans along $\mathbf{b}^*$ with narrow slits
(Fig. \ref{f-qscans}b).
These results show that each incommensurate satellite
$(q_1, 0.5, 0)$ actually is the superposition of two
satellites $(q_1, 0.5\pm\delta, 0)$ with $\delta=0.011$ at
$T = 35$ K.
In $\omega$-scans they are not resolved,
because of the relatively large width
of the reflections due to the less than optimal crystal quality.

The two-fold superstructure below $T_{c1}$ of TiOBr was determined
from the integrated intensities of the Bragg reflections measured
at $T = 17.5$ K.\cite{palatinusl2005a}
The result was in complete accordance with the two-fold
superstructure of TiOCl,\cite{shazm2005} and it thus
shows that the low temperature phase of TiOBr is a spin-Peierls
state like that in TiOCl.\cite{palatinusl2005a}

At $T = 35$ K the integrated intensities of the Bragg
reflections up to $(\sin(\theta)/\lambda)_{max} = 0.62$
{\AA}$^{-1}$ were measured by $\omega$-scans.
As noticed above, each measured satellite intensity corresponds
to the superposition of two reflections at
$({h+q_1}, {k+0.5\pm\delta}, l)$, or at
$({h-q_1}, {k+0.5\pm\delta}, l)$,
while the two groups are separated from each other.
Structure refinements were performed within the superspace
approach for incommensurate structures,
\cite{dewolff1981,vansmaalen1995} employing the computer
program {JANA2000}.\cite{petricekv2000}
The diffraction data have orthorhombic $mmm$ symmetry.
Refinements of the average structure against the intensities
of the main reflections showed that the room-temperature
structure in $Pmmn$ is also valid as the average structure at
35 K.
Cooling had negatively affected crystal quality, resulting
in Full Widths at Half Maximum (FWHMs) of the reflections
up to 0.06 degrees at 50 K.
A systematic variation of reflection widths could not be
observed between 50 K and 35 K.
However, the splitting of reflections due to a possible
lowering of symmetry is expected to be small, probably
less than the observed FWHM above the transition.
As a consequence, information on a possible
lowering of the symmetry towards monoclinic could not
be derived from the reflection profiles.


Without direct information on the symmetry, two possibilities
exist for the superspace group.
If the orthorhombic symmetry is preserved, the modulation
is two-dimensional with modulation wavevectors
\begin{eqnarray}
\label{e-ortho_qvectors}
\mathbf{q}^1\: &=& \: (0.075, 0.511, 0) \nonumber\\
\mathbf{q}^2\: &=& \: (-0.075, 0.511, 0)
\end{eqnarray}
at $T=35$ K.
The ${(3+2)}$-dimensional superspace group is
$Pmmn(\alpha,\beta,0)(-\alpha,\beta,0)000,000$.
The incommensurate modulation is described by one harmonic
for each of the two waves (Eq. \ref{e-ortho_qvectors}).
Because all atoms are in the mirror plane $m_a$, the
two harmonics are equivalent by symmetry and three of the
six independent components are zero.
This leads to modulation functions for each of the three
atoms Ti, O and Br given by three parameters only,
\begin{eqnarray}
\label{e-mod_function}
u_x(\bar{x}_{s4},\bar{x}_{s5}) &=& u_x^1 [
\sin(2\,\pi \bar{x}_{s4}) - \sin(2\,\pi \bar{x}_{s5}) ] \nonumber\\
u_y(\bar{x}_{s4},\bar{x}_{s5}) &=& u_y^1 [
\sin(2\,\pi \bar{x}_{s4}) + \sin(2\,\pi \bar{x}_{s5}) ]\\
u_z(\bar{x}_{s4},\bar{x}_{s5}) &=& u_z^1 [
\cos(2\,\pi \bar{x}_{s4}) + \cos(2\,\pi \bar{x}_{s5}) ] \nonumber
\end{eqnarray}
with $\bar{x}_{s4}=\mathbf{q}^1{\cdot}(\mathbf{x}+\mathbf{L})$ and
$\bar{x}_{s5}=\mathbf{q}^2{\cdot}(\mathbf{x}+\mathbf{L})$.
$\mathbf{x}$ is the average position of the atom in the unit cell,
and $\mathbf{L}$ enumerates the unit cells.

A reduction of the symmetry to monoclinic with unique axis
$\mathbf{a}$ gives the same model for the modulation functions
as was obtained in orthorhombic symmetry.
Therefore this possibility was not considered any further.
Alternatively, monoclinic symmetry with unique axis $\mathbf{c}$
corresponds to a one-dimensional modulation with wavevector
$\mathbf{q}=\mathbf{q}^1$, and with the ${(3+1)}$-dimensional
superspace group $P2/n(\alpha,\beta,0)00$.
The modulation now is a single wave, that can be obtained
from Eq. \ref{e-mod_function}
by removing all terms containing $\bar{x}_{s5}$.
The diffraction symmetry $mmm$ implies that the
crystal is twinned, if the symmetry is monoclinic,
with the first domain modulated by $\mathbf{q}^1$ and
the second domain modulated by $\mathbf{q}^2$.
Structure refinements, employing the nine independent
modulation parameters, gave excellent fits to the diffraction data
with equal $R$-factors for both monoclinic
symmetry (assuming twinning) and orthorhombic symmetry.
\footnote{502 measured reflections contain 159 observed
main reflections
and 149 observed superlattice reflections.
Structure refinements in monoclinic and orthorhombic symmetries
converged to models with the same reliability factors
$R_{all}=0.022$ with $R_{main}=0.018$ and $R_{sat}=0.080$.}
Furthermore, both models give rise to equal values for
the structure factors, so that they cannot be distinguished on the
basis of the diffraction data.

\section{Discussion}

Both structure models have equal values for the parameters,
except for
a scale factor of $\sqrt{2}$ between the modulation amplitudes,
that is explained by the different assumptions on twinning
(Table \ref{t-tiobr_modulation_displacements}).
The two models
share several features, that allow for conclusions on the
state of the intermediate phase.
The major amplitude is the displacement of Ti along $\mathbf{b}$.
This is similar to the displacements in the low-temperature
phase, and it suggests that antiferromagnetic interactions
between Ti atoms along $\mathbf{b}$ are important in the
incommensurate phase.
Displacements along $\mathbf{a}$ are small, in accordance
with the LT phase, where symmetry requires them to be zero.
Thus, the intermediate phase appears to be an incommensurate
version of the low-temperature
two-fold superstructure.
The incommensurability determines that the modulation amplitudes
as well as Ti--Ti distances assume all values between a
maximum and minimum.
Alternatively, it cannot be excluded that the modulation
is a block-wave,
that would correspond to domains with a two-fold
superstructure separated by domain walls
where the spin pairs would be broken.
Both the continuous modulation as well as the domain model
are in accordance with incomplete spin-pairing and with
the observed finite value of $\chi_m$ in the intermediate
phase.

\begin{table}
\caption{\label{t-tiobr_modulation_displacements}
Structural parameters for the three independent atoms
in the incommensurate structure of TiOBr at $T=35$ K.
Basic structure coordinates $\mathbf{x}=(x^0, y^0, z^0)$
are relative to the lattice parameters
$a=3.7849\, (9)$, $b=3.4685\, (7)$ and $c=8.500\, (2)$ {\AA}.
Modulation parameters are given in {\AA} (Eq. \ref{e-mod_function}).
Ti is at $(0, 0.5, z^0)$; O and Br are at $(0, 0, z^0)$.
}
\begin{ruledtabular}
\begin{tabular}{l c c c c}
Atom &   $z^0$       & $u_x^1$ ({\AA}) &
                                 $u_y^1$ ({\AA}) & $u_z^1$ ({\AA}) \\
\multicolumn{5}{l}{$Pmmn(\alpha,\beta,0)(-\alpha,\beta,0)000,000$} \\
Ti   &    0.11096 (6) &  -0.0043 (7)   &
          0.0262 (6)     & -0.0116 (6) \\
O    &   -0.05127 (23)&   0.0031 (33)  &
          0.0150 (20)    &  0.0160 (18) \\
Br   &    0.32947 (3) & -0.0033 (6)    &
         -0.0062 (4)     & -0.0148 (3)  \\
~\\
\multicolumn{5}{l}{$P2/n(\alpha,\beta,0)00$ ($\mathbf{c}$ unique)} \\
Ti   &  0.11096 (6) &  -0.0057 (11)  &
          0.0369 (9)     & -0.0165 (8)  \\
O    & -0.05126 (23)&   0.0050 (46)  &
          0.0213 (29)    &  0.0225 (26) \\
Br   &  0.32947 (3) &  -0.0049 (8)   &
         -0.0088 (5)     & -0.0210 (4)  \\
\end{tabular}
\end{ruledtabular}
\end{table}

\begin{figure}
{\includegraphics[width=4.0cm]{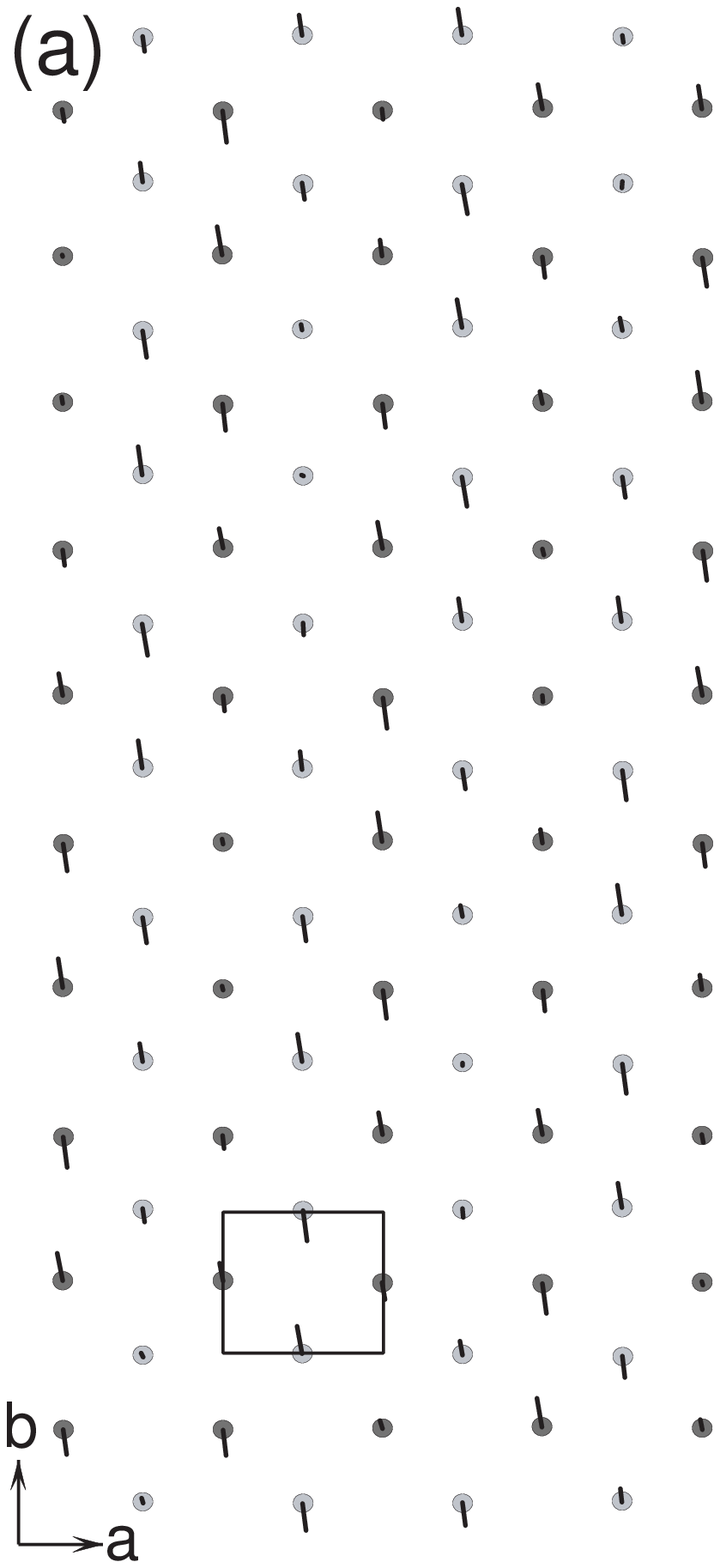}}\hfill
{\includegraphics[width=4.0cm]{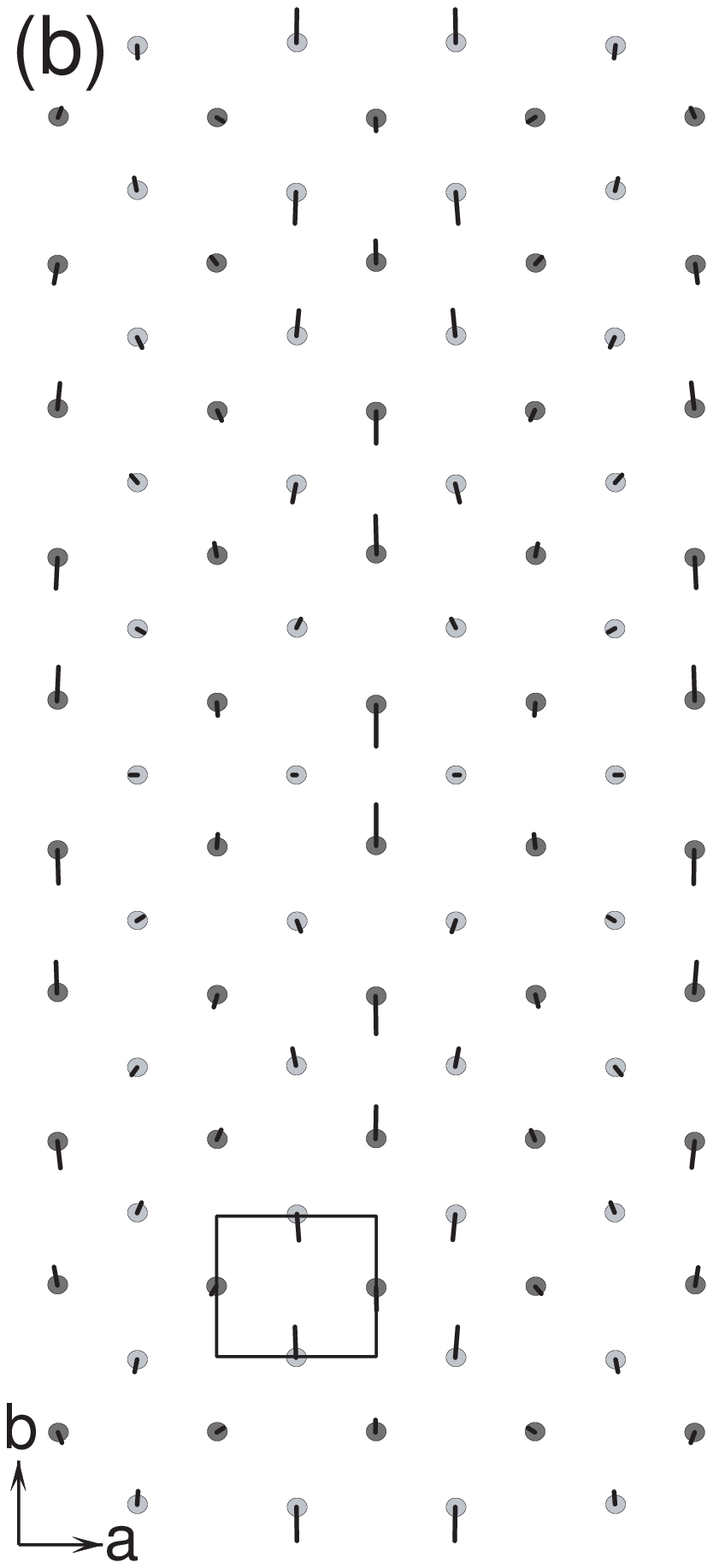}}
\caption{\label{f-displacements_supercell}
One double layer of Ti atoms parallel to the
$\mathbf{a},\mathbf{b}$ plane.
(a) For the 1D modulation in monoclinic symmetry.
(b) For the 2D modulation in orthorhombic symmetry.
Atoms are depicted at their basic positions, with shifts
towards their true position indicated by arrows with
a length of 20 times the true displacements.
Dark and light circles represent Ti atoms at
$-z_0$ and $z_0$, respectively.
The modulation was computed in a $5\times 11$ supercell
approximation.
}
\end{figure}

R\"uckamp \textit{et al.}\cite{ruckampr2005a} have proposed that
the incommensurate phase is the result of frustration between
intra- and interchain interactions, for which a finite
amplitude of the modulation along $\mathbf{a}$ is required.
The monoclinic model with a $1D$ modulation is in accordance
with this interpretation.
This model features displacements along $\mathbf{a}$ that are
small but vary in phase with the major displacement
along $\mathbf{b}$ (Eq. \ref{e-mod_function} and
Fig. \ref{f-displacements_supercell}a).

Interference of the two waves in the orthorhombic model
implies that the displacements along $\mathbf{a}$ and
$\mathbf{b}$ are out of phase
(Eq. \ref{e-mod_function} and Fig. \ref{f-displacements_supercell}b),
\textit{i.e} a large displacement along $\mathbf{b}$ corresponds
to zero displacement along $\mathbf{a}$.
This is exactly as can be expected for a competition between
intra- and interchain interactions,
the former dominating in regions with large $u_y$ displacements
and the latter being important in regions with small $u_y$.
The model of frustrated spin-Peierls interactions
could also be supported by this interpretation of the crystal
structure.\cite{ruckampr2005a}
Alternatively, the $2D$ nature of the modulation wave in orthorhombic
symmetry suggests that the incommensurability might
be the result of a competition between spin-Peierls interactions
on the chains and $2D$ magnetic interactions between neighboring
Ti atoms along $[0.5, 0.5, {\sim}0.25]$, resulting in a helical
magnetic structure coupled to the lattice distortion.
\cite{beynonrj1993}
The latter interpretation is in accordance with the
observed increase of the two-dimensional character
on increasing temperature,\cite{caimig2004a,lemmensp2004a}
and with the proposed
contributions of $d_{xz}$ and $d_{yz}$ symmetries to the
orbital of lowest energy and the increased admixture of
these orbitals at higher temperatures due to phonons.
\cite{gracol2004,pisanil2005,caimig2004a,lemmensp2004a,lemmensp2005}

The two symmetries lead to completely different properties of
the phase transition at $T_{c1}$.
Structurally, the orthorhombic model can be considered a precursor
to the spin-Peierls state, and the phase transition is a
lock-in transition.
On the other hand, the monoclinic model requires that the unique
axis switches from $\mathbf{c}$ to $\mathbf{a}$ at $T_{c1}$.
Accordingly, the domain structure of the material must change,
and a much more sluggish behavior of the transition,
\textit{e.g.} hysteresis, would be expected.

Previously, we have failed to observe incommensurate satellites
in the intermediate-temperature phase of {TiOCl}.
This is easily explained by the experimental method, that involved
scans along reciprocal lattice directions, and thus might
have missed possible satellites, especially if the
incommensurate components in {TiOCl} will be larger than
in {TiOBr}.

\section{Conclusions}

We have found that TiOBr exhibits a second-order
phase transition at $T_{c2} = 47.1 (4)$ K towards
an incommensurately modulated structure.
Available experimental data indicate that this structure
is either orthorhombic with a two-dimensional modulation
or monoclinic with a one-dimensional modulation.
Both structure models give rise to similar local structures.
Arguments have been given that favor or disfavor mechanisms
for the existence of the incommensurate phase, that involve
either frustrations between dimerized chains
\cite{ruckampr2005a} or two-dimensional magnetic interactions
\cite{beynonrj1993} in competition with one-dimensional
spin-Peierls interactions.
The equivalence of the low-temperature structures of TiOBr
and TiOCl shows that TiOBr is in a spin-Peierls
state below $T_{c1} = 26.8 \pm 0.3$ K.
\cite{shazm2005,palatinusl2005a}

\begin{acknowledgments}

Single crystals were grown by A. Suttner.
We thank W. Morgenroth for assistance with the
synchrotron experiment
at beam-line D3 of Hasylab at DESY (Hamburg, Germany).
Financial support by the German Science Foundation (DFG) is
gratefully acknowledged.
\end{acknowledgments}


\clearpage

\end{document}